\documentclass{PoS}
\usepackage{float}
\usepackage{amsmath,amssymb}
\usepackage{graphicx}
\usepackage{mathtools,slashed}
\usepackage{bm}
\usepackage{comment}
\usepackage{xcolor}
\usepackage{subfigure}
\usepackage{array}
\usepackage{multirow}
\usepackage{url}
\usepackage{color}
\usepackage[T1]{fontenc}
\usepackage[latin9]{inputenc}
\usepackage{graphicx}
\usepackage{esint}
\usepackage{atbegshi}
\usepackage{lipsum}

\newcommand{\W}[0]{{\mathcal{W}}}
\newcommand{\cW}[0]{\mathcal{W}}

\newcommand{\M}[0]{{\mathcal{M}}}
\newcommand{\Wtildf}[0]{{\mathcal{W}}_{L,{\rm df}}}

\newcommand{\w}[0]{w}

\title{Finite-volume matrix elements of two-body states}

\ShortTitle{Finite-volume formalism}

\author{\speaker{Alessandro Baroni}\\
Department of Physics and Astronomy University of South Carolina, 712 Main Street, Columbia, South Carolina  29208, USA\\
E-mail: \email{abaro008@odu.edu}}

\author{Ra\'ul A. Brice\~{n}o \\
Thomas Jefferson National Accelerator Facility, 12000 Jefferson Avenue, Newport News, Virginia 23606, USA\\
Department of Physics, Old Dominion University, Norfolk, Virginia 23529, USA \\
	E-mail: \email{rbriceno@jlab.org}}
	
\author{Maxwell T. Hansen\\
Theoretical Physics Department, CERN, 1211 Geneva 23, Switzerland\\
E-mail:\email{maxwell.hansen@cern.ch}}
	
\author{Felipe G. Ortega-Gama\\
Department of Physics, College of William and Mary, Williamsburg, Virginia 23187, USA\\
E-mail:\email{felortga@jlab.org}}

\abstract{In this talk, we present a framework for studying structural information of resonances and bound states coupling to two-hadron scattering states. This makes use of a recently proposed finite-volume formalism~\cite{Briceno:2015tza} to determine a class of observables that are experimentally inaccessible but can be accessed via lattice QCD. In particular, we shown that finite-volume two-body matrix elements with one current insertion can be directly related to scattering amplitudes coupling to the external current. For two-hadron systems with resonances or bound states, one can extract the corresponding form factors of these from the energy-dependence of the amplitudes. 
}

\FullConference{The 36th Annual International Symposium on Lattice Field Theory - LATTICE2018\\
		22-28 July, 2018\\
		Michigan State University, East Lansing, Michigan, USA.}

\begin{document}
\section{Introduction}
Lattice QCD (LQCD) calculations of multi-hadron systems are expected to play a crucial role in addressing a variety of questions about the nature of the strong interaction. An example of this is the study of resonance properties (for example of the $\rho$ meson) and in particular their form factors. Although the elastic form factors of resonances are experimentally inaccessible, they could provide important information of the structure and ultimately the nature of low-lying QCD states that are unstable. Another example, relevant for nuclear physics, is the study of electroweak processes in two-body systems (such as $n+p\rightarrow d+\gamma$ and the proton-proton fusion process driving our sun). Among other things, LQCD will be able to constrain and guide more standard methodologies used to study low-energy nuclear physics, such as effective field theories.

LQCD calculations of multi-hadron systems give finite-volume matrix elements that, in order to acquire physical meaning, must be connected to the corresponding infinite-volume amplitudes.  In some cases it is possible to derive a finite-volume formalism that maps the energies and matrix elements calculated on the lattice to physically-relevant, infinite-volume quantities. Early work in this direction was done by  L\"uscher~\cite{Luscher:1986pf, Luscher:1990ux}, who gave a relation between the finite-volume energy spectrum below the lowest multi-particle threshold and infinite-volume scattering amplitude. This framework for two-body systems has been generalized in a series of publications~\cite{Rummukainen:1995vs, Feng:2004ua, He:2005ey, Leskovec:2012gb,  Briceno:2012yi, Hansen:2012tf, Guo:2012hv, Briceno:2014oea} to particles with generic spin and to multiple channels. More recently it has been shown that is possible to derive (and implement) perturbative~\cite{Detmold:2014fpa} and all-orders~\cite{Lellouch:2000pv, Kim:2005gf, Christ:2005gi, Hansen:2012tf, Meyer:2011um, Feng:2014gba, Meyer:2012wk,  Bernard:2012bi, Briceno:2015csa, Briceno:2014uqa, Briceno:2015tza} relations between electroweak finite-volume matrix elements and the corresponding transition amplitudes.%
\footnote{For a recent review on these techniques and their implementation, we point the reader to Ref.~\cite{Briceno:2017max}}

In the following we discuss the model-independent and relativistic, finite-volume formalism developed in  Ref.~\cite{Briceno:2015tza} to study reactions of the type $\textbf 2 + J \to \textbf 2$ (with $\textbf 2$ denoting two-hadron states and $J$ a local current). We  explain how the formalism can be used to extract resonant form factors, and as a specific example we will consider the electromagnetic form factors of the $\rho$ meson. We further describe some of the crucial steps in the implementation of this framework, which is complicated in part by the appearance of a new kinematic function (an analog of the generalized zeta-function \cite{Luscher:1990ck} that arises in ${\bf 2}\rightarrow{\bf 2}$).

\section{Roadmap}
\begin{figure}[H]
	\begin{center}
	\includegraphics[width=6in]{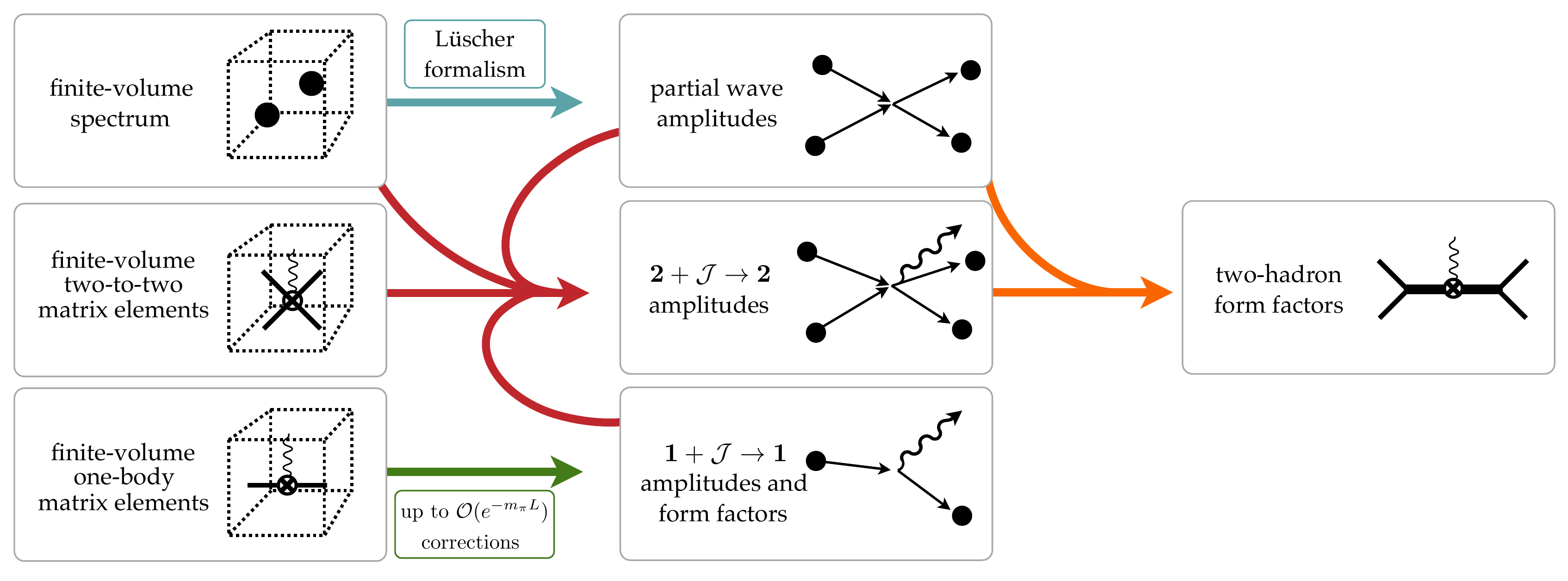}
	\caption{Scheme that shows the workflow necessary to extract the $\rho$ meson form factors from numerical LQCD calculations. See the text for a detailed description.
	 \label{fig:map}}
\end{center}
\end{figure}

The extraction of the electromagnetic form factors of the $\rho$ from numerical LQCD calculations requires a series of ingredients and a somewhat involved procedure, which is sketched in Fig.~\ref{fig:map}.  We recall that, since the $\rho$ is a resonance, its electromagnetic form factors can be defined by analytically continuing the initial and final $\pi \pi$ energies, appearing in the $\pi\pi\gamma^\star\rightarrow\pi\pi$ scattering amplitude, to the $\rho$ pole in the complex plane.%
\footnote{ 
For a recent implementation of this, we point the reader to Ref.~\cite{Briceno:2015dca}, where the transition $\pi\to\gamma^\star\rho$ form factor was obtained by analytically continuing the $\pi\to\gamma^\star\pi\pi$ amplitude onto the $\rho$-pole.
}
This is shown by the horizontal orange arrow in Fig.~\ref{fig:map}.
Therefore the main problem we are left with is how to obtain the $\pi\pi\gamma^\star\rightarrow\pi\pi $ from LQCD calculations, and to this aim we need three finite-volume ingredients and corresponding mappings that are summarized in the points below.
\begin{itemize}
	\item The finite-volume spectrum of two pions related to the infinite-volume $\pi\pi$ scattering amplitude by means of the well-established L\"uscher formalism.
	\item The finite-volume matrix element for one-to-one process with one insertion of the external current $J$, that is equal to the single pion form factor (up to a kinematic prefactor and  exponentially suppressed volume corrections).
	\item The finite-volume matrix element for two-to-two processes with one insertion of the external current $J$, that can be related to the corresponding infinite-volume scattering amplitude using the formalism derived in  Ref.~\cite{Briceno:2015tza} whose crucial ingredient is a new set of finite-volume functions. We note here that in order to perform this mapping, ingredients from the two previous points are needed as shown from the the curved red arrows in Fig.~\ref{fig:map}.
	\end{itemize}

Since the first two points are well-known, we will discuss our progress in the implementation of the third step.

\section{Covariant finite-volume formalism for ${\bf 2}+J\rightarrow {\bf 2}$}

We focus on the simplified case where the two incoming and two outgoing mesons have degenerate masses and the current insertion has one Lorentz index. The 4-vectors $P^\mu_i \equiv (E_i , \textbf P_i)$ and $P^\mu_f \equiv (E_f , \textbf P_f)$ denote respectively the 4-momentum in the finite-volume frame of the incoming and outgoing state. The energies in the center-of-momentum (c.m.) frame are
\begin{equation}
E_i^\star \equiv \sqrt{P_i^2} = \sqrt{E_i^2 - \textbf P_i^2} \,, \ \ \ \text{and} \ \ \ 
E_f^\star \equiv \sqrt{P_f^2} = \sqrt{E_f^2 - \textbf P_f^2} \,.
\end{equation}
Adopting the notation of Ref.~\cite{Kim:2005gf}, 
$\star$  denotes quantities in either the incoming or the outgoing c.m.~frame. 
Our goal is to extract the following amplitude
\begin{equation}
\cW_{\mu}(P_f, k'; P_i, k) \equiv \langle   P_f, k'; \text{out}    \vert   J_{\mu}(0)  \vert  P_i, k; \text{in}     \rangle_{\rm conn} \,,
\label{eq:mat:elem}
\end{equation}
where $k$ and $k^\prime$ are the on-shell four-momenta of one of the ingoing and outgoing particles respectively, and the ``${\rm conn}$'' subscript refers to the fact that only the fully connected piece is kept. The initial and final states in Eq.~(\ref{eq:mat:elem}) are standard two-particle asymptotic states and $J_{\mu}(0)$ is a local vector current insertion.  

This amplitude has radiative divergences, which do not contribute to the finite-volume matrix element. With this in mind, we define the divergence-free amplitude
\begin{align}
\label{eq:Wdf}
\W_{{\rm df};\mu} \equiv \W_{\mu} - 
i  \overline{\M}(P_f,k',k) \frac{i}{(P_f-k)^2 - m^2} \w_{\mu}
-
\w_{\mu}
\frac{i}{(P_i - k')^2 - m^2}
i \overline{\M^\dagger}(P_i, k',k)
\,,
\end{align} 
where $\overline{\M}$ is closely related to the full two-to-two scattering amplitude on the mass shell and $\w_{\mu}$ is the single-particle matrix element of $J_\mu$. (See Ref.~\cite{BBHO:2018} for more details.) 
 
Note that $\W_{\rm df}$ is defined by subtracting kinematic singularities from the full scattering amplitude for $\textbf 2+ J\rightarrow\textbf 2$. Here we see a major difference in this work as compared to Ref.~\cite{Briceno:2015tza}. In particular, the definition of the quantity $\W_{{\rm df};\mu}$ is fully Lorentz covariant. The two definitions are both valid, but the one used here simplifies the implementation of the formalism. 

We are now in a position to show the final formula that allows one to connect the finite-volume matrix elements with the corresponding infinite-volume quantities~\cite{Briceno:2015tza}
\begin{equation}
\Big |  \langle P_f, L \vert  J^{\mu}(0)  \vert  P_i, L \rangle \Big |^2 
=\frac{1}{L^6} {\rm{Tr}}\left[
\mathcal R( P_i) 
\Wtildf^{\mu}(P_i,P_f,L)
\mathcal R( P_f)  
\Wtildf^{\mu}(P_f,P_i,L)
\right] \,,
\label{eq:2to2}
\end{equation}
 where the left-hand side is the finite-volume matrix element.
The right-hand side involves a trace over the different possible angular momenta and the quantity $\mathcal R$ is 
a generalization of the Lellouch-L\"uscher factor \cite{Briceno:2015tza, Lellouch:2000pv, Briceno:2015csa, Briceno:2014uqa}. 

Finally, $\mathcal W_{L,\rm{df}}$ is related to the infinite-volume $\textbf 2 + J \to \textbf 2$ amplitude through 
 \begin{equation}
\label{eq:WtiltoWdf}
\mathcal{W}_{L,\rm{df}}^{\mu}(P_f,P_i,L)-\mathcal{W}_{\rm df}^{\mu}(s_i,s_f,Q^2) \equiv  
f_{\pi^+}(Q^2)  \mathcal  M (s_f)  \Big [ 
  (P_f + P_i)^\mu  G(P_f,P_i,L)  - 2   G^\mu(P_f,P_i,L)    \Big ]  \mathcal M(s_i) \, ,
\end{equation}
  where $s_i \equiv P_i^2$, $s_f \equiv P_f^2 $, $\mathcal M$ is the P-wave scattering amplitude, $Q^2=-(P_f-P_i)^2$, $f_{\pi^+}(Q^2)$ is the electromagnetic form factor the of the $\pi^+$,  $G$  and $G^\mu$ are new kinematic functions.%
  \footnote{In writing this expression, we have ignored contributions form the electromagnetic form-factor of the $\pi^0$, which vanishes at $Q^2=0$ but will have a small contribution for non-zero $Q^2$.} 
   In writing Eq.~(\ref{eq:WtiltoWdf}) we have suppressed indices associated the angular momenta of the initial and final states. As an example, we provide an explicit expression for $G^\mu$ when the initial and final states have zero total angular momenta,
  \begin{align}
{G}^{ \mu} (P_f,P_i,L)=
& \lim_{\alpha \to 0} \Big[ \frac{1}{L^3}\sum_{\mathbf{k}} -\int \! \frac{d^3 \textbf k}{(2\pi)^3}\Big]\,
\,
\frac{H(\alpha,\textbf{k}) }{2\sqrt{k^2+m^2}}
\frac{k^\mu}
{  (P_f-k)^2 - m^2 + i \epsilon  } 
\frac{1}
{  (P_i-k)^2 - m^2 + i \epsilon  }
 \, ,\label{eq:Gtot}
\end{align}
where $k^0=\sqrt{k^2+m^2}$ and $H(\alpha, \textbf{k})$ is an ultraviolet cut-off function, which is smooth in the kinematic window of interest, equal to one at each of the poles of the integrand, and equal to one in the limit that $\alpha\to0$. In Ref.~\cite{BBHO:2018} we provide explicit expressions for these functions for all angular momenta. 

The sum appearing in Eq.~(\ref{eq:Gtot}) can be efficiently calculated, and we have found the integral to be the most challenging part. In particular, standard numerical integration methods converge very slowly due to the singular structure of the integrand.
 We find it convenient  to write the integral as a sum of two parts: a singular four-dimensional piece and a three-dimensional {\it smooth} piece. In particular, it is possible to reduce the four-dimensional integral to a one-dimensional integral using standard techniques for the evaluation of one loop diagrams with three propagators. The {\it smooth} (and convergent) three-dimensional integral can be integrated with standard numerical routines. For the special case in which $P_i=P_f$, we are able to write the kinematic functions in terms of linear combinations of the generalized zeta-functions. We have performed various tests demonstrating that these two procedures agree when $P_i=P_f$. 

\begin{figure}[H]
	\begin{center}
		\includegraphics[width=6in]{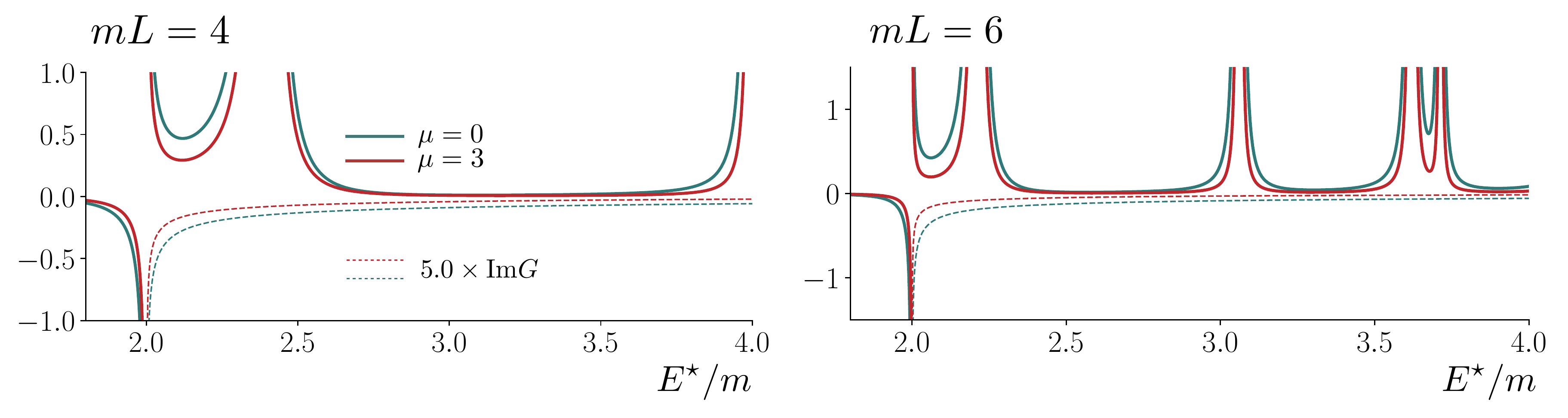}
		\caption{$G^\mu$ in units of $[m]^{-1}$ as a function of the center of mass energy for the case described in the text using two different values of the volume. The solid and dashed lines respectively correspond to the real and imaginary parts.  }
		\label{fig:GPiEQPf}
	\end{center}
\end{figure}

In Fig.~\ref{fig:GPiEQPf} we plot ${G}^{ \mu}$ for two volumes, when $P_i = P_f$, $\textbf{P}=2\pi[001]/L$, and the initial and final angular momenta are $(\ell_i,m_i)=(\ell_f,m_f)=(1,0)$. Similar to the zeta-functions present in the L\"uscher formalism, these functions are divergent when the energy of the system coincides with the free states. Unlike the zeta-functions, these have double poles. 

\section{Conclusions}
  In this talk we have presented a formalism for the study of $\textbf 2 + J \to \textbf 2$  reactions via lattice QCD. These amplitude may be used to extract structural information of bound states and resonances coupling to two-body states. We have developed efficient techniques for evaluating a new class of finite-volume functions that emerge in this formalism. 
  
  There are a several open challenges left to address in order to implement this framework, and these are currently under investigation. First, the amplitudes that appear are expressed in terms of the partial-wave basis, while the form factor of resonances/bound states are more naturally understood using Lorentz decomposition. Second, for the special case where one is interested in the extraction of the three electromagnetic form factors of the $\rho$ resonance, it will be necessary to constrain four Lorentz scalar amplitudes for real values of the energy of the initial and final $\pi\pi$ states. After analytic continuation to the $\rho$ pole, one linear combination of these amplitudes must vanish. Third, as with the vast majority of finite-volume observables, there is not a one-to-one mapping between matrix elements and scattering amplitudes. One possible solution here is to use well-motivated parametrizations to fit a large kinematic window of the matrix elements using Eq.~(\ref{eq:2to2}).  Similar techniques are currently being implemented in the analysis of coupled-channel spectra~\cite{Dudek:2016cru,Briceno:2017qmb}.  Ultimately, one would want to use amplitudes that satisfy dispersive relations. There has been recent development on this front for closely-related amplitudes, namely $\pi\gamma\to\pi\pi$~\cite{Hoferichter:2012pm, Hoferichter:2017ftn}, but more work is needed to accommodate $\pi\pi\gamma^\star\to\pi\pi$.

\section{Acknowledgments}
{We thank J. Dudek, R. Edwards, A. Jackura, A. S. Kronfeld, S. R. Sharpe, D. Pefkou, for useful discussions. The work of A.B. has been supported by the U.S. Department of Energy, Office of Science, Office of Nuclear Physics, under Award No. DE-SC0010300.

\bibliography{bibi}{}
\bibliographystyle{unsrt}

\end{document}